\documentclass[aps,prd,floats,preprint,tightenlines,nofootinbib]{revtex4}

\begin{document}
\preprint{ \hbox{hep-ph/0405199} }
\vspace*{3cm}

\title{ Family Unification on an Orbifold}
\author{
        Zhenyu Han\footnote{email address:  {\tt zhenyu.han@yale.edu}}
        and  Witold Skiba\footnote{email address:  {\tt witold.skiba@yale.edu}}}
\affiliation{        \small \sl  Department of Physics, Yale University,
                New Haven, CT  06520\vspace{2.5cm} }

\begin{abstract}
We construct a family-unified model on
a $Z_2\times Z_2$ orbifold in five dimensions.
The model is based on a supersymmetric $SU(7)$
gauge theory. The gauge group is broken by orbifold
boundary conditions to a product of grand unified
$SU(5)$ and $SU(2)\times U(1)$ flavor symmetry.
The structure of Yukawa matrices is generated
by an interplay between spontaneous breaking of flavor symmetry
and geometric factors arising due to field localization
in the extra dimension.
\end{abstract}

\maketitle

\newpage

 \section{Introduction}
 The success of gauge coupling
unification in the Minimal Supersymmetric Standard Model (MSSM)
 suggests there exists a Grand Unified Theory (GUT). In GUTs,
 one generation of fermions can be incorporated in one or more
 representations of a simple GUT gauge group. However,
 GUTs do not explain why there are three
different families and do not shed any light on the pattern of
 the observed fermion mass spectrum and mixing angles.
An immediate idea is to also assign a symmetry group for the
generations, the so-called flavor group. Theories incorporating this idea
can be divided into two broad categories. First, theories in which the flavor
group and the GUT group are orthogonal. Second, theories that
unify the flavor group and the GUT group into a larger simple group.
The second approach is known as family unification~\cite{family}.
Within the first category, realistic theories with continuous flavor groups $SU(3)$,
$SU(2)\times U(1)$, $U(1)$ as well as several discrete groups have been considered.
Family unified models push the unification idea a step further and are esthetically
more attractive. However, four-dimensional models of family unification
 usually suffer from the problem of mirror families, see Ref.~\cite{Babu2002} for a review.

In the last few years the notion of symmetry breaking by orbifold
boundary conditions in extra dimensions has been revitalized.
Orbifold breaking has been used to address various problems
ranging from electroweak symmetry breaking~\cite{orbESB}, supersymmetry
breaking~\cite{orbSUSY} to GUT model building~\cite{orb2-3,orbGUT}. For example, in
Ref.~\cite{orb2-3} orbifold compactification is used to break
the $SU(5)$ GUT group to the Standard Model group and solve the
doublet-triplet splitting problem that is difficult to overcome in 4D
models. One of the reasons we use orbifold boundary conditions is to
give large masses to mirror fermions, as noticed for example in Ref.~\cite{Babu2002}.

In this article, we construct a family-unified model in 5 dimensions.
To maintain the unification of the gauge couplings our model
incorporates supersymmetry. The fifth dimension is compactified
and we impose orbifold boundary conditions on all fields propagating
in the fifth dimension. The role of the boundary conditions is threefold.
We use the orbifold breaking to get rid of mirror families, break family-unified
gauge group to a product of GUT and flavor symmetry, and also
reduce the amount of supersymmetry to ${\mathcal N}=1$ in 4D.\
Given that the quarks of the third generation are a lot heavier than the quarks
of the first two generations,  it seems natural that the
light families form a doublet, while the third family a singlet
under the flavor group. Consequently, models  using $SU(2)\times U(1)$ flavor
 symmetry  are quite successful in reproducing the mass spectrum~\cite{SU2xU1}.
We embed the flavor $SU(2)\times U(1)$ and $SU(5)$ GUT group in an
$SU(7)$ family unified gauge group. A similar setup was studied in
Ref.~\cite{Hwang2002}, where an $SU(7)$ family unified model was considered.
However, in Ref.~\cite{Hwang2002} the GUT group is flipped $SU(5)$ and the
emphasis is on the doublet-triplet splitting problem. A number of authors discussed
flavor in extra dimensions, see Ref.~\cite{EDflavor} and references within.

The $SU(7)$ gauge group is broken by the boundary conditions to
$SU(5)\times SU(2)\times U(1)$. Both the $SU(5)$ and the flavor groups are
broken further by expectation values of Higgs fields. The pattern of Yukawa
matrices is generated by both spontaneous breaking of the
flavor group and geometric factors due to field localization. Some
of the fields in our model propagate in the bulk, while others
are localized at the orbifold fixed points. Therefore, wavefunction
overlap suppresses certain couplings with respect to others.
All of the flavor physics takes place at very high energy scales,
comparable to the GUT scale. Supersymmetry breaking
terms are of order the electroweak scale and are irrelevant
for the discussion of flavor. We will not discuss the breaking
of ${\mathcal N}=1$ supersymmetry in any detail since this is not the focus
of this paper. Any standard mechanism of communicating supersymmetry
breaking in a flavor-diagonal manner could be incorporated into
our model. Standard gauge mediation~\cite{gm} could operate if SUSY breaking
and messenger fields are localized at an orbifold fixed point.  By extending the model to
one more dimension one could create an appropriate setup for either
anomaly mediation~\cite{am} or gaugino mediation~\cite{ggm} of supersymmetry breaking.

In the next section, we describe the field content and
interactions needed to produce Yukawa matrices.
We summarize our results in Section~\ref{sec:summary}.
The details concerning numerical determination
of the high-energy parameters from the data are
presented in Appendix~\ref{app:fitting}.

\section{The model}
\label{sec:model}
Our model is based on a supersymmetric field theory
in five dimensions. The fifth dimension is compactified on a
$(Z_2\times Z_2)$ orbifold. We parameterize the
fifth dimension, described by coordinate $y$, as an interval with  $ y\in [0,\frac{\pi R}{2}]$.
This interval can be thought of as obtained from a circle $[0,2\pi R]$ by identifying
points related by reflections around two perpendicular axes.
Under these reflections,  $y\sim-y$ and $ y \sim\pi-y $ such that the
circle is equivalent to the $ y\in [0,\frac{\pi R}{2}]$ interval.
We denote these reflections as $P$ and $P'$, respectively.

 An arbitrary bulk field configuration can be decomposed into
the eigenstates of the reflections $P$ and $P'$.
Since $P^2=P'^2=1$ the eigenvalues must be $\pm1$.
Of course, the eigenstates of the reflections have
either the Dirichlet or Neumann boundary conditions
at the end points of the interval.  The Kaluza-Klein (KK) decomposition
of a bulk field $\phi(x^\mu,y)$ into four dimensional mass
eigenstates can be classified according to the two parities:
\begin{eqnarray}
\label{eq:KK}
 \phi_{++}(x^\mu,y)&=&\sum_{n=0}^\infty \frac{1}{\sqrt{2^{\delta _{n0}}\pi
   R}}\phi_{++}^{2n}(x^\mu)\cos \frac{2ny}{R}, \label{KK} \\
 \phi_{+-}(x^\mu,y)&=&\sum_{n=0}^\infty \frac{1}{\sqrt{\pi
   R}}\phi_{+-}^{2n+1}(x^\mu)\cos \frac{(2n+1)y}{R}, \nonumber \\
  \phi_{-+}(x^\mu,y)&=&\sum_{n=0}^\infty \frac{1}{\sqrt{\pi
    R}}\phi_{-+}^{2n+1}(x^\mu)\sin \frac{(2n+1)y}{R}, \nonumber \\
 \phi_{--}(x^\mu,y)&=&\sum_{n=0}^\infty \frac{1}{\sqrt{\pi
    R}}\phi_{--}^{2n+2}(x^\mu)\sin \frac{(2n+2)y}{R}, \nonumber \end{eqnarray}
where $x^\mu$ is the four dimensional coordinate and the subscripts
refer to the parities under the $P$ and $P'$ reflections.
The five dimensional Lagrangian has simple dependence on $y$ when
the fields are expressed in terms of KK states. The integral over
the fifth dimension can be performed explicitly.
One obtains then a four-dimensional Lagrangian describing
a KK tower of four dimensional fields. The KK states specified in Eq.~(\ref{eq:KK})
have masses  $\frac{2n}{R}$, $\frac{2n+1}{R}$, $\frac{2n+1}{R}$, and $\frac{2n+2}{R}$,
respectively. The only massless 4D field is $\phi_{++}^0(x^\mu)$.

It turns out that the compactification scale in our model
will be comparable to the GUT scale. The massive states will
therefore be too heavy to correspond to observable states. The
fields of the MSSM will come from the zero modes of the KK
decomposition, as well as from brane fields localized at the endpoints
of the interval.

\subsection{Fields and interactions}
\label{sec:fields}
We now begin to describe our model in detail.
The 5D bulk theory is an ${\mathcal N}=1$ SUSY theory with an
$SU(7)$ gauge group. Such a theory has 8 supercharges
and corresponds to ${\mathcal N}=2$ SUSY in four dimensions.
However, the boundary conditions preserve only 4 supercharges,
so that below the compactification scale the theory is a four dimensional
${\mathcal N}=1$ theory.

There is an arbitrary choice of how the reflection symmetry is represented
in the space of gauge transformations.  We choose the action of the two
parities on the fundamental representation of the $SU(7)$ group to be
$P=diag\{1,1,1,1,1,1,1\}$ and $P'=diag\{-1,-1,-1,-1,-1,1,1\}$. Consequently,
an arbitrary tensor representation of $SU(7)$, $\phi_{kl\ldots}^{ij\ldots}$,
transforms as
\begin{eqnarray}
  \phi_{kl\ldots}^{ij...}(-y) &=& \eta_\phi
  P^i_{i'}P^j_{j'}P^{k'}_{k}P^{l'}_{l}\ldots\phi^{i'j'\ldots}_{k'l'\ldots}(y), \nonumber \\
  \phi_{kl\ldots}^{ij...}(\pi-y)&=&\eta'_\phi
  P'^i_{i'}P'^j_{j'}P'^{k'}_{k}P'^{l'}_{l}\ldots\phi^{i'j'\ldots}_{k'l'\ldots}(y), \nonumber
\end{eqnarray}
under the two parity transformations, where $\eta_\phi,\eta_\phi'=\pm1$ are the overall, "internal",
 parity eigenvalues. For a free field the parities can be chosen arbitrarily. Interaction
terms correlate the parities of different fields. For example, the invariance of the supersymmetric
Lagrangian imposes relations between parities of different components of superfields.

 The 5D gauge multiplet contains a vector $A_M$, two
 gauginos $\lambda_1$, $\lambda_2$, and a real scalar $\Sigma$, all of which
transform in the adjoint representation of $SU(7)$. We use the upper case Latin letters to denote
5D Lorentz indices, and the lower case Greek letters to denote 4D indices.
The 5D SUSY Lagrangian is invariant under the reflections if
\begin{equation}
\eta_{_{A_\mu}}=-\eta_{_{A_5}}=-\eta_{_\Sigma}, \quad
\eta_{_{\lambda_1}}=-\eta_{_{\lambda_2}},
\end{equation}
as well as an identical set of relations for $P'$. We choose
$\eta_{_{A_\mu}}=\eta'{_{A_\mu}}=\eta_{_{\lambda_1}}=\eta'_{_{\lambda_1}}=1.$

Upon compactification, the first reflection breaks the ${\mathcal N}=1$ 5D
SUSY to ${\mathcal N}=1$ 4D SUSY since  both $A_5$ and $\lambda_2$  obtain
large masses. Meanwhile, $A_\mu$ and $\lambda_1$ contain the zero modes
that transform exactly as the 4D ${\mathcal N}=1$ vector multiplet.
Since we do not embed the parity
 transformations into the R symmetry ${\mathcal N}=1$ supersymmetry in 4D is preserved.
 The second reflection breaks the gauge group from $SU(7)$ to its
 $ SU(5)\times SU(2)\times U(1)$ subgroup.
More precisely, on the
brane located at $y=\frac{\pi R}{2}$ the gauge group is broken, while in the
bulk and on the brane $y=0$ the full symmetry remains. Besides the
gauge multiplets, we put the 5D hypermultiplets in the bulk.
Under the $SU(7)$ symmetry, the hypermultiplets  transform as
$\textbf{1}+\textbf{7}+\textbf{35}+\bf{\overline{21}} $.\footnote{The $SU(7)$ field content
coincides with an $SO(14)$ spinor $\textbf{64}$ when the spinor is written in the
SU(7) basis.  This suggests that our model may be embedded in a larger
symmetry group.} A hypermultiplet corresponds to two 4D chiral
superfields with opposite parities $\{\Psi, \Psi^c\}$:
\begin{eqnarray}
\eta_\Psi&=&-\eta_{\Psi^c},  \\
\eta'_{\Psi}&=&-\eta'_{\Psi^c}.
\end{eqnarray}
We choose $\eta_\Psi=\eta'_\Psi=1$ so that all the massless fields
come from $\Psi$. These massless fields can be expressed in terms of  representations of the unbroken gauge groups $ SU(5)\times SU(2)\times U(1)$. These representations are
$T^a ({\bf 10},{\bf 2})_{-1}$, $F^a  ({\bf\bar{5}},{\bf 2})_3$, $S^a ({\bf 1},{\bf 2})_{-5}$,  and a neutral field $({\bf 1},{\bf 1})_0$, where $a=1,2$ is the SU(2) index. This set of 4D massless fields is free of gauge anomalies. In addition,
there is no 5D anomaly either in the bulk or on the
branes \cite{anomaly}. We interpret the $SU(5)$ as the GUT group
and the $SU(2)\times U(1)$ as a flavor group.  The zero modes
$({\bf 10},{\bf  2})_{-1}$ and $({\bf \bar{5}},{\bf  2})_3$ are chosen to be the light two
families of fermions with their superpartners and $({\bf 1},{\bf 2})_{-5}$
might be the right handed neutrinos with their superpartners.

On the asymmetric brane we add all other superfields that
are necessary to complete the MSSM spectrum and break the
GUT and flavor symmetries. Since the $SU(7)$ gauge
symmetry is broken on this brane, the localized fields do not need to form
complete $SU(7)$ multiplets. First,
we choose the third family to be SU(2) singlets:
$T_3 ({\bf 10},{\bf 1})_0$, $ F_3 ({\bf \bar{5}},{\bf 1})_0$.
Second, the SU(5) symmetry is broken by the conventional Higgses:
$\Sigma ({\bf 24},{\bf 1})_0$, $H ({\bf 5},{\bf 1})_0$, $\overline{H} ({\bf \bar{5}},{\bf 1})_0$,
$K ({\bf 45},{\bf 1})_0$, and  $\overline{K} ( {\bf \overline{45}},{\bf 1})_0$.
We need to include the  ${\bf 45}$ and ${\bf \overline{45}}$ representations  to avoid the
undesirable relations $m_\mu\simeq m_s$ and $m_e\simeq m_d$ at the GUT
scale. In addition, we introduce the following fields to break
the flavor  $SU(2)\times U(1)$ symmetry:
$\phi_{T,a} ({\bf 1},{\bf 2})_1$, $\overline{\phi}_T^a ({\bf 1},{\bf 2})_{-1}$,
$\phi_{Fa} ({\bf 1},{\bf 2})_{-3}$,
 $\overline{\phi}_F^a ({\bf 1},{\bf 2})_3$, $\psi ({\bf 1},{\bf 1})_{-2}$, and
 $\overline{\psi} ({\bf 1},{\bf 1})_2$.
We will refer to these fields as "flavons". We summarize the light field content in Table~1.
\begin{table}[hbt]
\begin{tabular}{|c|c|c|}
\hline
   & bulk fileds &  brane fields \\ \hline
matter &   $T^a ({\bf 10},{\bf 2})_{-1}$, $F^a  ({\bf \bar{5}},{\bf 2})_3$, $S^a ({\bf 1},{\bf 2})_{-5}$
     & $T_3 ({\bf 10},{\bf 1})_0$, $ F_3 ({\bf \bar{5}},{\bf 1})_0$ \\ \hline
Higgs &   & $\Sigma ({\bf 24},{\bf 1})_0$, $H ({\bf 5},{\bf 1})_0$,
                    $\overline{H} ({\bf \bar{5}},{\bf 1})_0$ \\
  fields & &  $K ({\bf 45},{\bf 1})_0$,   $\overline{K} ({\bf  \overline{45}},{\bf 1})_0$ \\ \hline
flavons & & $\phi_{T,a} ({\bf 1},{\bf 2})_1$, $\overline{\phi}_T^a ({\bf 1},{\bf 2})_{-1}$,
$\psi ({\bf 1},{\bf 1})_{-2}$, $\overline{\psi} ({\bf 1},{\bf 1})_2$ \\
& &  $\phi_{F,a} ({\bf 1},{\bf 2})_{-3}$, $\overline{\phi}_F^a ({\bf 1},{\bf 2})_3$ \\ \hline
\end{tabular}
\caption{Light chiral superfields and their $SU(5)\times SU(2)\times U(1)$ charges.}
\end{table}

Having listed the field content of our model, we now show how to obtain appropriate Yukawa
couplings.  Since our model is five-dimensional, the underlying field theory is nonrenormalizable
and has a cutoff $\Lambda$ that is roughly two orders of magnitude larger than the
compactification scale. To specify the Yukawa couplings we write the superpotential
in terms of four-dimensional fields that is the brane fields and
the zero modes of the bulk fields:
\begin{eqnarray}
  W&=&T_3T_3H+T_3F_3\bar{H}+ \frac{1}{\Lambda}
             \left[  T_3T\phi_TH+F_3T\phi_T\bar{H}+T_3 F\phi_F\bar{H}
                     + T _3 T \phi_T K \right. \nonumber\\
      &&  \left. +F_3T\phi_T\bar{K} +T_3F\phi_F\bar{K}+TF\psi\bar{H}+TF\psi\bar{K}\right]+\
                \frac{1}{\Lambda^2} \left[ T\phi_TT\phi_TH\right. \nonumber \\
      && \left. +T\phi_TF\phi_F\bar{H}  + T\phi_TT\phi_TK+T\phi_TF\phi_F\bar{K}
             +TT\Sigma H\bar\psi +TT\Sigma K\bar\psi \right].
 \label{eq:WYukawa}
\end{eqnarray}
This superpotential is valid for a 4D theory below the compactification scale,
so it is easy to keep track of dimensions of operators. We have explicitly indicated the
$\frac{1}{\Lambda}$ suppression of dimension five and dimension six terms.
Different terms in Eq.~(\ref{eq:WYukawa}) are not related by any symmetries,
so each term comes with a different coefficient. We have omitted the coefficients
of operators for now. We will define and determine these coefficients in Appendix~\ref{app:fitting}.

In Eq.~(\ref{eq:WYukawa}) we included almost all dimension five and six terms allowed
by the gauge symmetries. We have omitted the couplings of $K$ and $\bar{K}$ to the third family
fields $T_3$ and $F_3$. Also, the flavons $\overline{\phi}_F$, $\overline{\phi}_T$ do not
appear in Eq.~(\ref{eq:WYukawa}). Since the superpotential is not renormalized it is technically
natural to exclude certain terms. However, we can assign global symmetries
to our fields such that  the unwanted terms in Eq.~(\ref{eq:WYukawa})
are prohibited. Such symmetries also prevent $\overline{\phi}_F$ and $\overline{\phi}_T$
from appearing in the higher order terms, like $\frac{1}{\Lambda^3}$,  $\frac{1}{\Lambda^4}$, etc.
For any term, the addition of the $SU(5)$ adjoint $\Sigma$
is allowed by gauge symmetries. As we will show later, $\langle \Sigma \rangle / \Lambda$ is small. Thus, we also omitted terms with powers of $\Sigma$ whenever they would modify a lower order
term that is more important.

For the sake of clarity, we have also omitted
geometric suppression factors in Eq.~(\ref{eq:WYukawa}). These factors are important for
the structure of Yukawa matrices and are written explicitly in Sec.~\ref{sec:Yukawas}.
Such factors arise when bulk fields couple to brane fields because bulk fields propagate
in 5D and their overlap with brane fields is small.  A geometric factor is
$\rho=\frac{1}{\sqrt{\Lambda R \pi/2}}$.  For a given term, the number of powers of
$\rho$  suppressing the term is equal to the number of bulk fields present in the term~\cite{Hall2001}.

The ${\bf 45}$ representation, $K$, and its conjugate $\overline{K}$ contain one $SU(2)$ doublet
each. Together with the doublets coming from $H$ and $\overline{H}$
there would be four light doublets. We assume that one linear combination of
doublets from $H$ and $K$ gets a large mass and the orthogonal linear combination
remains light. The same thing takes place for $\overline{H}$ and $\overline{K}$.
We outline how to realize this in Appendix~\ref{app:mixing}.
We denote the light mass eigenstates as
$h_{u,d}$ and the heavy ones as $h'_{u,d}$. In terms of these mass eigenstates
\begin{eqnarray}
  H_u=\sin\alpha h_u+\cos\alpha h'_u, &&  K_u=\cos\alpha h_u-\sin\alpha h'_u, \\
  H_d=\sin\gamma h_d+\cos\gamma h'_d, && K_d=\cos\gamma h_d-\sin\gamma h'_d,
\end{eqnarray}
where $H_{u,d}$ and $K_{u,d}$ represent the $SU(2)$ doublet components
of the corresponding fields.
The mixing angles $\alpha$ and $\gamma$ are free parameters. For convenience, we define
$v=\cot\gamma$ and  $v'=\cot\alpha$.

 We assume that the flavons and $\Sigma$ get  D-flat, SUSY-preserving, VEVs:
\begin{equation}
\label{eq:flavonvev}
   \frac{\langle\phi_{T}\rangle}{\Lambda}=\frac{\langle\overline{\phi}_T\rangle}{\Lambda}
            =\left(\begin{array}{c} 0 \\ \epsilon \end{array} \right),\quad
   \frac{\langle\phi_{F}\rangle}{\Lambda}=\frac{\langle{\overline{\phi}_F}\rangle}{\Lambda}
      =\left(\begin{array}{c} 0 \\ \epsilon' \end{array} \right),
    \quad\frac{\langle\psi\rangle}{\Lambda}=\frac{\langle\overline\psi\rangle}{\Lambda}=\sigma,
\end{equation}
\begin{equation}
  \label{eq:sigmavev}
   \quad\frac{\langle\Sigma\rangle}{\Lambda}={\rm Diag}\{-\frac23\delta,-\frac23\delta,-\frac23\delta,\delta,\delta\}.
\end{equation}
Supersymmetry is only broken by weak-scale soft masses. We do not specify the superpotential
that produces these VEVs as it is not essential for our discussion, but it would not be difficult to do so.

\subsection{Yukawa matrices}
\label{sec:Yukawas}
Using Eqs.~(\ref{eq:WYukawa}) through (\ref{eq:sigmavev}) it is straightforward to
write the Yukawa matrices in terms of the MSSM superfields. These couplings arise
at the GUT scale after $\langle \Sigma \rangle$ breaks $SU(5)$ to the Standard Model.
We denote the Yukawa matrices as $Y_u$,
$Y_d$ and $Y_l$ for the up quarks, down quarks, and charged leptons, respectively.
We do not consider the neutrino mass matrices or CP violating phases in this article.
The Yukawa matrices are given by
\begin{eqnarray}
  Y_u&\sim&\left(\begin{array}{ccc}0&
                                \rho^2\sigma\delta(1+v)&0\\
                               -\rho^2\sigma\delta(1+v)&\rho^2\epsilon^2(1+v)
                                &\frac12\rho\epsilon(1+v)\\0&\frac12\rho\epsilon(1+v)&1
                              \end{array}\right) \sin\gamma, \label{eq:yu1} \\
  Y_d&\sim&\left(\begin{array}{ccc}0&
                                  \rho^2\sigma(1+v')&0\\-\rho^2\sigma(1+v')&\rho^2\epsilon\epsilon'(1+v')
                                  &\rho\epsilon(1+v')\\0&\rho\epsilon'(1+v')&1
                               \end{array}\right) \sin\alpha,  \label{eq:yd1} \\
 Y_l&\sim&\left(\begin{array}{ccc}0&
                                \rho^2\sigma(1-3v')&0\\-\rho^2\sigma(1-3v')&\rho^2\epsilon\epsilon'(1-3v')
                                &\rho\epsilon(1-3v')\\0&\rho\epsilon'(1-3v')&1
                            \end{array}\right)\sin\alpha. \label{eq:yl1}
\end{eqnarray}
We indicated the matrices with the proportionality sign because in the interest of clarity we omitted
arbitrary coefficients that are also missing in Eq.~(\ref{eq:WYukawa}). Compare Eqs.~(\ref{eq:yu2})-(\ref{eq:yl2})
in Appendix~\ref{app:fitting} that contain the full set of coefficients.
The factors of $\frac{1}{2}$ that appear in the (2,3) and
(3,2) elements of $Y_u$ arise because the terms
$T_3T\phi_TH$ and $T_3T\phi_TK$ contribute to both elements.

We determine the magnitudes of the elements of the Yukawa matrices at the
GUT scale by using the renormalization group equations for these matrices
and comparing them with the masses and the CKM angles at the weak scale.
The fitting procedure is described in Appendix~\ref{app:fitting}.
As we will see the experimental data can be fitted quite accurately.
Before we present the results let us make several comments about
the matrices (\ref{eq:yu1})-(\ref{eq:yl1}).

If $\epsilon\approx \epsilon'$  the structure of the quark Yukawa couplings
is very similar to the 4 texture zero symmetric quark mass matrices
discussed in the literature. See Ref.~\cite{Fritzsch2000}
for a review. (Since the matrices discussed in  Ref.~\cite{Fritzsch2000}
are symmetric the off-diagonal zeros are the same element and counted as one zero.)
The only difference is that our matrices are antisymmetric
in the (1,2) and (2,1) indices~\cite{SU2xU1,Berezhiani:1998vn}.
The matrix for the leptons is similar to that for the
down quarks. The zeros in the Yukawa matrices  (\ref{eq:yu1})-(\ref{eq:yl1})
are exact provided that $\overline{\phi}_F$ and $\overline{\phi}_T$
are absent in the superpotential in Eq.~(\ref{eq:WYukawa}).

As discussed in Ref.~\cite{fritzsch2003}, this kind of matrices can
give us the approximate relations
\begin{eqnarray}
   |V_{us}|  &\approx  &\left|\sqrt{\frac{m_u}{m_c}}-e^{i\phi}\sqrt{\frac{m_d}{m_s}}\right|,  \\
  \left|  \frac{V_{td}}{V_{ts}} \right|  &\approx&   \sqrt{\frac{m_d}{m_s}},
\end{eqnarray}
where $\phi$ is a CP violating phase defined in Ref.~\cite{fritzsch2003}.
Of course, our matrices have only real elements, so the phase in the first relation is absent.
In order to avoid the undesirable relation
\begin{equation}
    \left| \frac{V_{ub}}{V_{cb}}\right| \approx \sqrt{\frac{m_s}{m_b}}
\end{equation}
 the (2,3) and (3,2) elements of the Yukawa matrices should be sufficiently
 large--much larger than $m_s/m_b$ and $m_c/m_t$ for d-Yukawa
 matrix and u-Yukawa matrix, respectively.

The (2,3) and (3,2) elements of $Y_d$ and $Y_l$ are too large to be neglected
compared to the (2,2) and (3,3) diagonal elements. The off-diagonal elements
are different for $Y_d$ and $Y_l$: they are $\rho \epsilon (1+v')$ and $\rho \epsilon (1-3 v')$,
respectively. Thus, the off-diagonal elements affect the largest eigenvalue of the matrix
differently for the bottom Yukawa and the $\tau$ Yukawa. Therefore, the bottom-$\tau$
unification is not exact and the $b$ and $\tau$ masses can be fitted accurately.
Similar observation was made in Ref.~\cite{Barr:2002mw}.

It is not possible to uniquely determine all the parameters
in Eqs.~(\ref{eq:yu1})-(\ref{eq:yl1}) like $\epsilon$, $v$, etc.\
because there are arbitrary coefficients $a_i$ in front of every term,
see Appendix~\ref{app:fitting}. Only certain combinations
of the coefficients $a_i$ and other parameters appear in the
Yukawa matrices. We would like  all coefficients $a_i$ to be close to one since
they are dimensionless couplings. We get the values of $a_i$
to be close to one by choosing the remaining parameters as
follows
\begin{equation}
  \label{eq:values}
  \rho\epsilon = \frac{1}{30},\quad\rho\epsilon' = \frac{3}{40},
  \quad\rho^2 \sigma = \frac{3}{2000},\quad\delta=\frac{1}{20}, \quad
  v' = \frac{5}{3} , \quad v=\frac{2}{3}.
\end{equation}
We assume that $\rho\approx0.1$ and infer the VEVs of flavons:
$\epsilon\approx0.33$, $\epsilon' \approx 0.75$, and $\sigma\approx0.15$.
These VEVs are smaller than the cutoff $\Lambda$, but $\epsilon'$
is quite close to 1. Together with $\delta\approx0.05$ and the GUT scale
$M_{GUT}\sim2.8\times10^{16}{\mbox{GeV}}$, we get
$\Lambda\sim5.6\times10^{17}\mbox{GeV}$ and
$\frac1R\sim8.8\times10^{15}\mbox{GeV}$.  These numbers give a resonable
separation of  the cutoff, the GUT, and the
compactification scales~\cite{Hebecker2002}. However, the VEVs of the flavons
are sufficiently close to the cutoff scale that higher dimensional operators
may play an important role in the generation of Yukawa couplings in our model.
The values of parameters in Eq.~(\ref{eq:values}) correspond to $\tan \beta=47$,
but we could make similar choices for other values of $\tan \beta$.

What is interesting is that the flavons' VEVs: $\epsilon$, $\epsilon'$, $\sigma$
are of the same order. This is very different from many 4D models where the
flavons usually obtain hierarchical VEVs in order to produce hierarchy in the
Yukawa matrices. The geometric suppression factor
does contribute in our model to generating small ratios.

\section{Summary}
\label{sec:summary}
The underlying theory for our model is a 5D SUSY theory with an
$SU(7)$ gauge group. Compactification of the fifth dimension on
a $Z_2\times Z_2$ orbifold breaks SUSY to ${\mathcal N}=1$ in 4D
as well as breaks $SU(7)$ to GUT $SU(5)$ times flavor $SU(2)\times U(1)$.
The compactification scale is very close to the GUT scale, it is just a factor
of three smaller than the GUT scale. Thus, our model is an ordinary
SUSY GUT almost all the way to the GUT scale.
In addition to symmetry breaking by boundary conditions we
introduce two types of Higgs fields. First, standard Higgs fields that
 break GUT symmetry down to the Standard Model and give masses to
 the quarks and leptons. Second, flavon Higgs fields whose role
 is to completely break the flavor symmetry. The flavor symmetry is broken
 close to the GUT scale.

Bulk multiplets contain zero modes corresponding to the two
lightest families that transform as a doublet under flavor $SU(2)$.
The third family is a singlet under the flavor symmetry and it is
localized at one of the orbifold fixed points. The $SU(7)$ gauge symmetry
is not preserved at the fixed point where the third family is localized.
Therefore, the third family does not  come from a complete
$SU(7)$ multiplet and is a flavor singlet.
As far as the flavor symmetry and the light fields are concerned our model
is very similar to the 4D model described in Ref.~\cite{SU2xU1}.

Our main goal was constructing a realistic pattern of Yukawa
matrices at the GUT scale. We were only concerned with the
quark and charged lepton sectors and completely neglected
the neutrino sector. The Yukawa couplings come from
the superpotential in Eq.~(\ref{eq:WYukawa}), which we chose to resemble
the "four zeros" texture described in Ref.~\cite{Fritzsch2000}.
The resulting Yukawa matrices, omitting a number of dimensionless
constants of order one, are given in Eqs.~(\ref{eq:yu1}), (\ref{eq:yd1}), and (\ref{eq:yl1}).

The orders of magnitude of different elements of the Yukawa matrices
are governed by three different effects. The first effect is the geometry
of our model. The couplings that involve both localized fields and bulk fields
are suppressed due to small wavefunction overlap between such fields.
Second, the $SU(2)\times U(1)$ flavor symmetry is broken by three
different flavons and their conjugates. Among the three flavons
there are two $SU(2)$ doublets and one singlets. All flavons
are charged under the $U(1)$. We do not count separately the
conjugates of the flavons because the VEVs of flavons with the conjugate
quantum numbers are identical to maintain SUSY above the
weak scale. Third, the up and down sectors are distinguished by
the mixing of the Higgs doublets that come both from the ${\bf 5}$
and the ${\bf 45}$. The light up and down Higgs doublets come
from different linear combinations of ${\bf 5}$ and ${\bf 45}$.
Of course, any value of $\tan \beta$ other than 1 also differentiates
the up and down sectors.

Our model has too many free parameters to be predictive.
What we accomplished, however, is generating the Yukawa
matrices in terms of a few small parameters: flavon VEVs, defined
in Eqs.~(\ref{eq:flavonvev}) and (\ref{eq:values}),  and
the geometric suppression factor. By matching to the
observed fermion mass spectrum and quark mixing
angles we determined the 13 nonzero parameters in
the Yuklawa matrices, see Eqs.~(\ref{eq:yu})-(\ref{eq:yl}).
We chose the undetermined parameters such that the
dimensionless couplings are close to one.

What is interesting is that given a few arbitrary choices
all dimensionless coefficients are of order one.
Moreover, many of the coefficients listed in Eq.~(\ref{eq:as})
are very close to one. All the large ratios are determined in terms
of the geometric suppression factor and a
few flavon VEVs that are of the same order of magnitude.
Obviously, a more fundamental
and predictive structure of flavor is still missing. However, it is conceivable
that the flavor could be generated from an interplay between geometry and
flavor symmetries.

\section*{Acknowledgments}
We thank M. Piai for discussions and  comments on the manuscript.
This research was supported in part by the US Department of Energy
under grant  DE-FG02-92ER-40704. WS is also supported in part by the DOE OJI program.

\appendix
\section{Mixing of Higgs doublets}
\label{app:mixing}
We briefly comment on the mixing of the Higgs doublets coming from the
${\bf 45}$ and ${\bf 5}$ representations and their conjugates. As we explained
in Sec.~\ref{sec:model}, we need the ${\bf 45}$ and ${\bf \overline{45}}$
representations to avoid the equality of the lepton-down quark Yukawa couplings
in the two light families. The problem is similar to $SO(10)$ unification,
where one needs to introduce larger Higgs representation
in addition to the ${\bf 10}$-dimensional Higgs to incorporate realistic Yukawa
couplings. The additional Higgs fields, for
example ${\bf \overline{126}}$, would produce too many light doublets.
A simple solution was presented in Ref.~\cite{lee1995}.
Similar solution works in the $SU(5)$ case and
we outline it here for completeness.

We supplement the Higgs fields $H$,
$\bar{H}$, $K$, and $\bar{K}$ introduced already by  another pair of ${\bf 45}$ and
${\bf \overline{ 45}}$ Higgs fields. Let us refer to the new fields as $K_1$ and $\overline{K}_1$.
We assume that the superpotential for these Higgs fields is given by
\begin{equation}
 W_{Higgs}=\mu H \overline{H} + H \Sigma\overline{H}+H \Sigma\overline{K}_1+
 \overline{H} \Sigma K_1+M_1 K_1\overline{K}+M_2 K\overline{K}_1.
\end{equation}
In the equation above $\Sigma$ is the $SU(5)$ adjoint field
that develops an $SU(5)$ breaking VEV given by Eq.~(\ref{eq:sigmavev}) and
$\langle \Sigma \rangle \propto \delta \Lambda$. $M_1$ and $M_2$ are arbitrary
mass parameters that are comparable to the GUT scale.
We also assume that $\mu\approx-\delta \Lambda$ so that the
$SU(2)$ doublets in $H$ and $\overline{H}$ are light. The mass matrix for the
doublets arising from $H$, $\bar{H}$, $K$, $\bar{K}$, $K_1$, and $\bar{K_1}$ has the following structure
\begin{equation}
  {\mathcal M}= \left( \begin {array}{c}H_d\\ K_{1d}\\K_d\end{array} \right)^T
  \left( \begin{array}{ccc}  0& \delta\Lambda & 0 \\
                                           \delta\Lambda&0&M_2 \\
                                           0 & M_1 & 0
        \end{array}  \right)
  \left( \begin {array}{c}H_u\\ K_{1u}\\K_u\end{array} \right),
\end{equation}
where we assumed that $\mu+\delta \Lambda=0$.
The light eigenvalues of this mass matrix are two doublets
\begin{eqnarray}
   h_u  &=& \cos\alpha\, H_u+\sin\alpha \, K_u, \\
   h_d  &=& \cos\gamma\, H_d+\sin\gamma\, K_d,
\end{eqnarray}
where $\tan\alpha=\delta\Lambda/M_2$ and $\tan\gamma=
\delta\Lambda/M_1$. There is no reason to assume that  $M_1$ and $M_2$ are equal,
so the mixing angles of  the up and down Higgs doublets are, in general,  different.
Clearly, the remaining Higgs doublets have masses of order the
unification scale and so do other components of $K$, $\overline{K}$, $K_1$, and
$\overline{K}_1$.

\section{Fitting to the data}
\label{app:fitting}

We now describe our procedure for determining the Yukawa matrices
at the GUT scale.  As we mentioned in Sec.~\ref{sec:fields}, there are no relations between the various terms in
Eq.~(\ref{eq:WYukawa}) since different terms are not related by symmetries to one another.
Therefore, one needs to include arbitrary coefficients of
order one in front of every term. After including the missing coefficients
Eq.~(\ref{eq:WYukawa}) becomes:
\begin{eqnarray}
  W&=& a_1 T_3T_3H+ a_2 T_3F_3\bar{H}+ \frac{1}{\Lambda}
             \left[ a_3 T_3T\phi_TH+ a_4 F_3T\phi_T\bar{H}+ a_5 T_3 F\phi_F\bar{H}
                     + a_6 T _3 T \phi_T K \right. \nonumber\\
      &&  \left. +a_7 F_3T\phi_T\bar{K} +a_8 T_3F\phi_F\bar{K}+ a_9 TF\psi\bar{H}+a_{10}TF\psi\bar{K}\right]+
                \frac{1}{\Lambda^2} \left[ a_{11}T\phi_TT\phi_TH  \right.\\
      && \left.  +a_{12} T\phi_TF\phi_F\bar{H} + a_{13} T\phi_TT\phi_TK+a_{14} T\phi_TF\phi_F\bar{K}
             +a_{15} TT\Sigma H\bar\psi + a_{16} TT\Sigma K\bar\psi \right].  \nonumber
\end{eqnarray}

The corresponding Yukawa matrices are then
\begin{eqnarray}
  Y_u&=& \left(\begin{array}{ccc}0&
                           \rho^2\sigma\delta(a_{15}+a_{16}v)&0\\
                           -\rho^2\sigma\delta(a_{15}+a_{16}v)&\rho^2\epsilon^2(a_{11}+a_{13}v)
                           &\frac{\rho\epsilon}{2}(a_3+a_6v)\\
                           0&\frac{\rho\epsilon}{2}(a_3+a_6v)&a_1
                      \end{array}\right) \sin\gamma, \label{eq:yu2}\\
   Y_d&=&\left(\begin{array}{ccc}0&
                          \rho^2\sigma(a_9+a_{10}v')&0\\
                          -\rho^2\sigma(a_9+a_{10}v')&\rho^2\epsilon\epsilon'(a_{12}+a_{14}v')
                          &\rho\epsilon(a_4+a_7v')\\
                          0&\rho\epsilon'(a_5+a_8v')&a_2
                     \end{array}\right)\sin\alpha, \label{eq:yd2}\\
  Y_l&=&\left(\begin{array}{ccc}0&
                               \rho^2\sigma(a_9-3a_{10}v')&0\\
                               -\rho^2\sigma(a_9-3a_{10}v')&\rho^2\epsilon\epsilon'(a_{12}-3a_{14}v')
                               &\rho\epsilon(a_4-3a_7v')\\
                               0&\rho\epsilon'(a_5-3a_8v')&a_2
                          \end{array}\right) \sin\alpha. \label{eq:yl2}
\end{eqnarray}
The Yukawa matrices are defined in terms of 16 coefficients $a_i$, 3 flavon VEVs, $\Sigma$ VEV,
geometric factor $\rho$, and two Higgs mixing angles: a total of 23 parameters.
However, several of our parameters only appear in particular combinations, which
allows us to eliminate the "unobservable" combinations:
\begin{eqnarray}
  Y_u&=&\left(\begin{array}{ccc}0&c_1&0\\-c_1&c_2&c_3\\0&c_3&1\end{array}\right)\eta,\label{eq:yu}\\
  Y_d&=&\left(\begin{array}{ccc}0&c_4&0\\-c_4&c_5&c_6\\0&c_7&1\end{array}\right)\zeta,\label{eq:yd}\\
  Y_l&=&\left(\begin{array}{ccc}0&c_8&0\\-c_8&c_9&c_{10}\\0&c_{11}&1\end{array}\right)\zeta. \label{eq:yl}
\end{eqnarray}
We are left with 13 parameters: $c_1$ through $c_{11}$, $\eta$, and $\zeta$. The experimental data, gives nine masses and three real angles in the CKM matrix. Including complex phases
in our Yukawa matrices would introduce too many free parametres, so we omit the phases.
If we performed a fit with the phases present the values of the real parameters
might change slightly, but such change would not affect the structure of the Yukawa matrices.

To obtain the GUT scale values we use the following
fermion parameters and the gauge couplings at  the scale
$M_Z$~\cite{fusaoka} as inputs:
 $$\alpha_1 = 0.016829, \alpha_2 = 0.033493, \alpha_3 = 0.118,$$
$$ m_u = 2.33\pm0.435~{\rm MeV}, m_c = 0.677\pm 0.0585~{\rm GeV},
    m_t = 181\pm13~{\rm GeV},$$
$$m_d = 4.36\pm1.13~{\rm MeV}, m_s = 72\pm 23~{\rm MeV}, m_b = 3.00\pm0.1~{\rm GeV},$$
$$ m_e =486.84727\pm0.00014~{\rm keV}, m_\mu = 102.75138\pm0.00033~{\rm MeV}, m_\tau
   = 1.74669\pm0.000285~{\rm GeV},$$
$$V_{us} = 0.2205\pm 0.0018, V_{cb} = 0.0373\pm0.0018,|V_{ub}/V_{cb}| = 0.08\pm0.02. $$
Given the structure of Yukawa matrices described in Eqs.~(\ref{eq:yu})-(\ref{eq:yl})
at the GUT scale, we use the one loop renormalization group equations in the MSSM~\cite{Martin1994} to compare with  the weak scale data. We set the GUT scale to be $M_{GUT}=2.80\times10^{16}$ where the three gauge coupling constants unify.
The one-loop running
of the gauge couplings does not involve the Yukawa couplings, so
the gauge couplings are determined at all scales  before fitting the Yukawa matrices.
We neglect the fact that the first few KK modes appear below $M_{GUT}$
since $R M_{GUT} \approx 3.2$. Since the logarithm of $R M_{GUT}$
is small we can neglect the effects of the KK modes below $M_{GUT}$
and we use 4D RGE equations.

In practice, we numerically evaluate the RGE equations from the GUT scale
down to the weak scale. We then compare the results of the RGE running with the data
and evaluate the $\chi^2$ using the experimental errors.
The errors are severely underestimated this way because threshold corrections
and two-loop effects are much larger than the experimental uncertainty of the lepton masses.
However, since we have more parameters than the number of inputs we are able
to get a good fit.  For example, we
present the numerical fit for $\tan\beta=47$ below.
\begin{eqnarray}
Y_u&=&\left(\begin{array}{ccc}0&-0.0001050&0\\0.0001050&0.005335&0.05848\\0&0.05848&1\end{array}\right)1.053,\\
Y_d&=&\left(\begin{array}{ccc}0&0.004744&0\\-0.004744&0.006898&0.1009\\0&0.2205&1\end{array}\right)0.4597,\\
Y_l&=&\left(\begin{array}{ccc}0&-0.003774&0\\0.003774&-0.007916&-0.1526\\0&-0.3161&1\end{array}\right)0.4597.
\end{eqnarray}
The total $\chi^2$ for this fit is 4.654. The $\chi^2$ is dominated by
the errors from $m_d$ and $m_\tau$, but none of the two masses with the poorest fit
deviates by more than $1.5\sigma$ from the experimental value.

To extract the physical parameters $\rho,\sigma$ and so on, we
need to remember that the coefficients $a_1, \ldots,a_{16}$ are
close to 1. A choice of parameters is given in Eq.~(\ref{eq:values})
and the corresponding coefficients are
\begin{eqnarray}
\label{eq:as}
&a_1=1.266,\quad a_2=0.8935, \quad a_3=3,\quad a_4=1.006,\quad
a_5=1.029,\quad a_6=2.162,&\nonumber\\ & a_7=1.019,\quad
a_8=0.9590,\quad a_9=1.557,\quad a_{10}=0.7611,\quad a_{11}=3,
\quad a_{12}=1.141,&\nonumber \\ &a_{13}=4.618,\quad
a_{14}=0.7942, \quad a_{15}=1,\quad a_{16}=1.160.&
\end{eqnarray}
Note that the coefficients $a_3$, $a_{11}$ and $a_{15}$ in $Y_u$ are set by hand.
There are too many free parameters to be uniquely determined from the 13 parameters
in Eqs.~(\ref{eq:yu})-(\ref{eq:yl}), so we need to arbitrarily choose some of them.

Most coefficients $a_i$ are very close to 1 and certainly none of the coefficients deviate from
one by an order of magnitude.
There are many small additional contributions to this result that we neglected,
for example higher-dimensional operators, threshold corrections, higher-loop effects.
Since the coefficients $a_i$ are so close to multiples of 1 it is possible
that the deviations could be accounted for by higher-order effects we neglected.
This suggests that there could be a simple set of hidden symmetries
responsible for this result. It certainly would be an exciting possibility.


\end{document}